\documentclass[letter]{aa} 
\usepackage{graphicx}
\usepackage{flushend}

\usepackage{natbib,twoopt}
\usepackage[colorlinks,linkcolor=blue,urlcolor=blue,citecolor=black]{hyperref} 
\bibpunct{(}{)}{;}{a}{}{,}             
\makeatletter
  \newcommandtwoopt{\citeads}[3][][]{\href{http://ui.adsabs.harvard.edu/abs/#3}%
    {\def\hyper@linkstart##1##2{}%
     \let\hyper@linkend\@empty\citealp[#1][#2]{#3}}}
  \newcommandtwoopt{\citepads}[3][][]{\href{http://ui.adsabs.harvard.edu/abs/#3}%
    {\def\hyper@linkstart##1##2{}%
     \let\hyper@linkend\@empty\citep[#1][#2]{#3}}}
  \newcommandtwoopt{\citetads}[3][][]{\href{http://ui.adsabs.harvard.edu/abs/#3}%
    {\def\hyper@linkstart##1##2{}%
     \let\hyper@linkend\@empty\citet[#1][#2]{#3}}}
   \newcommandtwoopt{\citeauthorads}[3][][]%
    {\href{http://ui.adsabs.harvard.edu/abs/#3}
    {\def\hyper@linkstart##1##2{}%
     \let\hyper@linkend\@empty\citeauthor[#1][#2]{#3}}}
  \newcommandtwoopt{\citeyearads}[3][][]%
    {\href{http://ui.adsabs.harvard.edu/abs/#3}
    {\def\hyper@linkstart##1##2{}%
     \let\hyper@linkend\@empty\citeyear[#1][#2]{#3}}}
  \renewcommand*\aa@pageof{, page \thepage{} of \pageref*{LastPage}} 
\makeatother
\usepackage[varg]{txfonts}
\usepackage{pdflscape} 

\begin{document}

\title{Blue extreme disk-runaway stars with \textit{Gaia} EDR3}

\author{Andreas~Irrgang\inst{\ref{remeis}}
        \and
        Markus~Dimpel\inst{\ref{remeis}}
        \and
        Ulrich~Heber\inst{\ref{remeis}}
        \and
        Roberto~Raddi\inst{\ref{catalunya}}
       }

\institute{
Dr.~Karl~Remeis-Observatory \& ECAP, Astronomical Institute, Friedrich-Alexander University Erlangen-Nuremberg (FAU), Sternwartstr.~7, 96049 Bamberg, Germany\label{remeis}\\
\email{andreas.irrgang@fau.de}
\and
Universitat Polit\`ecnica de Catalunya, Departament de F\'isica, c/ Esteve Terrades 5, 08860 Castelldefels, Spain\label{catalunya}
}
\date{Received 21 December 2020 / Accepted 22 January 2021}

\abstract{
Since the discovery of hypervelocity stars in 2005, it has been widely believed that only the disruption of a binary system by a supermassive black hole at the Galactic center (GC), that is, the so-called Hills mechanism, is capable of accelerating stars to beyond the Galactic escape velocity. In the meantime, however, driven by the {\it Gaia} space mission, there is mounting evidence that many of the most extreme high-velocity early-type stars at high Galactic latitudes do originate in the Galactic disk and not in the GC. Moreover, the ejection velocities of these extreme disk-runaway stars exceed the predicted limits of the classical scenarios for the production of runaway stars. Based on proper motions from the {\it Gaia} early data release 3 and on recent and new spectrophotometric distances, we studied the kinematics of 30 such extreme disk-runaway stars, allowing us to deduce their spatial origins in and their ejection velocities from the Galactic disk with unprecedented precision. Only three stars in the sample have past trajectories that are consistent with an origin in the GC, most notably S5-HVS\,1, which is the most extreme object in the sample by far. All other program stars are shown to be disk runaways with ejection velocities that sharply contrast at least with classical ejection scenarios. They include HVS\,5 and HVS\,6, which are both gravitationally unbound to the Milky Way. While most stars originate from within a galactocentric radius of 15\,kpc, which corresponds to the observed extent of the spiral arms, a group of five stars stems from radii of about 21--29\,kpc. This indicates a possible link to outer Galactic rings and a potential origin from infalling satellite galaxies.
}


\maketitle
\section{\label{sect:introduction}Introduction}
Increasingly more early-type main-sequence (MS) stars in the Galactic halo have been reported to move at very high space velocities (see, e.g., \citeads{2019A&A...628L...5I}; \citeads{2020MNRAS.491.2465K}; \citeads{2020arXiv201108862R}, and references therein), and some may even exceed their local Galactic escape velocity (\citeads{2015ARA&A..53...15B}; \citeads{2018MNRAS.479.2789B}, and references therein). These stars are believed to have been born in the star-forming regions of the Milky Way, that is, inside the Galactic disk, and then been ejected from their birth environment. They are therefore generally referred to as runaway stars. Two classical ejection mechanisms are typically considered, the binary supernova scenario (BSS, \citeads{1961BAN....15..265B}) and the dynamical ejection scenario (DES, \citeads{1967BOTT....4...86P}). In the BSS, the runaway star is the former companion of a massive star that underwent a core-collapse supernova. The corresponding explosion disrupted the binary and released the secondary with almost its orbital velocity. In the DES, the runaway star is ejected from young clusters and associations through close stellar encounters. These dynamical interactions mostly occur during the initial relaxation phase of the star cluster and are most efficient between two close binaries.

These two classical scenarios can hardly explain the acceleration of stars to velocities beyond $\sim400$\,km\,s$^{-1}$ (see \citeads{2018A&A...620A..48I}; \citeads{2020MNRAS.497.5344E}, and references therein). Consequently, in order to produce what is commonly called a hypervelocity star (HVS), that is, a star moving faster than its local Galactic escape velocity, the so-called Hills mechanism \citepads{1988Natur.331..687H} was invoked when the first HVSs were discovered (\citeads{2005ApJ...622L..33B}; \citeads{2005A&A...444L..61H}; \citeads{2005ApJ...634L.181E}). \citetads{1988Natur.331..687H} showed that the tidal forces of a supermassive black hole could disrupt nearby binaries, with one component being captured in a tight orbit while the other is ejected at a velocity (thousands of km\,s$^{-1}$) that is well sufficient to overcome the gravitational attraction of the Galaxy. A systematic radial-velocity survey led to the discovery of 21 late B-type stars that qualified as candidate HVSs \citepads{2014ApJ...787...89B}. However, because astrometric data of sufficient precision and accuracy were lacking, it was not possible at that time to verify whether these 21 objects were indeed Hills stars, that is, objects ejected by the supermassive black hole at the Galactic center (GC). Nevertheless, the Hills mechanism was assumed to be the most likely explanation of the very high space motions of the HVS candidates.

While not useful for the HVS candidates, the quality of astrometric measurements in the pre-{\it Gaia} era at least allowed more nearby runaway stars to be studied. Based on proper motions from the {\it Hipparcos} mission, \citetads{2008A&A...483L..21H} concluded that the massive B-type star HD\,271791 is gravitationally unbound to the Milky Way, although its does not stem from the GC but from the outer rim of the Galactic disk. In order to distinguish these stars from HVSs, \citetads{2008ApJ...684L.103P} coined the term hyper-runaway star. In addition to the spatial origin, the ejection velocity is another crucial quantity for characterizing the nature of runaway stars. In a comprehensive study, \citetads{2011MNRAS.411.2596S} derived ejection velocities for a compilation of 96 B-type MS runaway stars. A high-velocity population of 11 stars with ejection velocities between 400--500\,km\,s$^{-1}$ was identified. Unfortunately, the respective uncertainties were large because of the currently poor quality of the underlying ground-based and {\it Hipparcos} proper motions.

The unprecedented precision of the {\it Gaia} second data release (DR2, \citeads{2016A&A...595A...1G}, \citeads{2018A&A...616A...1G}) became a game changer. \citetads{2018MNRAS.479.2789B} revisited the more than 500 mostly late-type HVS candidates proposed before and found most of them to be bound. Only for fewer than 100 of them did the probability to be unbound exceed 1\%, and only for 41 does it exceed 50\%. \citetads{2018A&A...620A..48I} and \citetads{2020A&A...637A..53K} reanalyzed 40 of the 42 most extreme stars from the MMT-HVS survey \citepads{2014ApJ...787...89B}, including most of the 21 HVS candidates mentioned above, and ruled out the Hills mechanism in the GC for almost all of the 18 targets that did have a relatively well-constrained spatial origin. \citetads{2019ApJ...873..116H} and \citetads{2020MNRAS.491.2465K} came to the same conclusion for the four HVS candidates from the LAMOST survey (\citeads{2014ApJ...785L..23Z}; \citeads{2017ApJ...847L...9H}; \citeads{2018AJ....156...87L}), and \citetads{2019A&A...628L...5I} showed that the relatively bright B-type MS star PG\,1610$+$062 was ejected with a velocity of no less than $550\pm40$\,km\,s$^{-1}$. However, overwhelming observational evidence for the operation of the Hills mechanism has also been found by \citetads{2020MNRAS.491.2465K}. They discovered that the A-type star S5-HVS\,1 travels at a Galactic rest-frame velocity of $1755\pm50$\,km\,s$^{-1}$ and most likely originates in the GC.

Nevertheless, there is a clear trend that most of the original HVS candidates are found to be runaway stars from the Galactic disk instead of Hills stars when the precision of the underlying astrometry is sufficient to distinguish between these two options. This is an interesting result because the ejection velocities of these discarded Hills stars still exceed the predicted limits of the classical ejection mechanisms (\citeads{2009ApJ...706..925B}; \citeads{2012ApJ...751..133P}; \citeads{2014ApJ...793..122K}; \citeads{2015MNRAS.448L...6T}; \citeads{2020MNRAS.497.5344E}), which depend on stellar mass. For instance, \citetads{2015MNRAS.448L...6T} reported that the maximum ejection velocity in the BSS is 540\,km\,s$^{-1}$ for a $3.5\,M_\odot$ star, while it is only 320\,km\,s$^{-1}$ for a $10\,M_\odot$ star. Consequently, the high observed ejection velocities of disk runaways call for an as yet mostly neglected mechanism possibly involving very massive stars (\citeads{2009MNRAS.395L..85G}; \citeads{2009MNRAS.396..570G}) or intermediate-mass black holes \citepads{2019MNRAS.489.4543F}, see \citetads{2018A&A...620A..48I} for a more detailed discussion. With proper motions of significantly improved accuracy and precision from the {\it Gaia} early data release 3 (EDR3, \citeads{2020arXiv201201533G}; \citeads{2020arXiv201203380L}), it is now possible to determine whether this trend continues.
\section{Target selection}
Our initial sample comprised all 40 stars of the MMT-HVS survey that were revised by \citetads{2020A&A...637A..53K}, together with 30 runaway stars from the collection of \citetads{2011MNRAS.411.2596S} for which we were able to obtain spectra. This covers the majority of objects with high ejection velocities in that compilation. This group was complemented by the prototype hyper-runaway star HD\,271791 \citepads{2008A&A...483L..21H}, the potential hyper-runaway stars SDSS\,J013655.91$+$242546.0 (J0136$+$2425 for short, \citeads{2009A&A...507L..37T}) and HIP\,60350 \citepads{2010ApJ...711..138I}, the extreme disk-runaway star PG\,1610+062 \citepads{2019A&A...628L...5I}, the four candidate HVSs from the LAMOST survey (\citeads{2014ApJ...785L..23Z}; \citeads{2017ApJ...847L...9H}; \citeads{2018AJ....156...87L}), and the probable Hills star S5-HVS\,1 \citepads{2020MNRAS.491.2465K}. Based on proper motions from {\it Gaia} EDR3, we then carried out spectroscopic and kinematic analyses (see Sects.~\ref{sect:analysis} and \ref{sect:kinematics} for details) of all members of this initial sample to filter out those with both a high ejection velocity and a relatively well-constrained origin within the Galactic disk. In order to account for the fact that massive stars are typically ejected at lower velocity, we chose a mass-dependent threshold for the deduced $1\sigma$ upper limit of the ejection velocity, that is, 400\,km\,s$^{-1}$ or 320\,km\,s$^{-1}$ for stars with masses below or above $5\,M_\odot$, respectively. The first cut applies to almost all stars from the MMT-HVS survey, while most of the others fall into the second category. The chosen thresholds roughly represent the limits for the classical ejection scenarios (see, e.g., \citeads{2015MNRAS.448L...6T}; \citeads{2018A&A...620A..48I}, and references therein). A disk origin was granted when the $1\sigma$ lower limit of the inferred galactocentric plane-crossing radius was below 25\,kpc (motivated by \citeads{2015ApJ...801..105X}), while visual inspection was used to judge whether the origin was sufficiently well constrained. These criteria left us with 14 stars from the MMT-HVS survey, 7 stars from the \citetads{2011MNRAS.411.2596S} sample, all 4 stars from the LAMOST survey, and the 5 individual targets mentioned above, yielding a final sample of 30 program stars, the names of which are listed in Table~\ref{table:vejection}.
\begin{table}
\small
\centering
\renewcommand{\arraystretch}{1.07}
\caption{\label{table:vejection}Ejection velocity, galactocentric radius at plane intersection, and stellar mass of the program stars.}
\begin{tabular}{lrrrrr}
\hline\hline
Object & \# & $\varv_{\mathrm{ej}}$ & $r_\mathrm{p}$ & $M$ & F \\
& & (km\,s$^{-1}$) & (kpc) & ($M_\odot$) \\
\hline
                S5-HVS\,1 &  1 &          $1810^{+60}_{-60}$ &     $0.19^{+0.64}_{-0.09}$ & $2.35^{+0.06}_{-0.06}$ & C \\
                     B537 &  2 &           $750^{+60}_{-60}$ &        $3.4^{+3.3}_{-1.9}$ & $3.73^{+0.26}_{-0.17}$ & D \\
B576\,\tablefootmark{(a)} &  3 &           $710^{+70}_{-40}$ &        $0.8^{+0.9}_{-0.5}$ &       $0.5 \pm \ldots$ & C \\
                     B598 &  4 &           $640^{+70}_{-50}$ &        $1.0^{+1.0}_{-0.7}$ & $2.40^{+0.05}_{-0.05}$ & C \\
                   HVS\,5 &  5 &           $627^{+34}_{-24}$ &        $9.5^{+1.1}_{-1.2}$ & $3.40^{+0.10}_{-0.10}$ & D \\
                   HVS\,6 &  6 &           $623^{+27}_{-26}$ &             $25^{+6}_{-6}$ & $3.04^{+0.09}_{-0.09}$ & R \\
                     B481 &  7 &           $612^{+24}_{-21}$ & $29^{+\phantom{0}7}_{-10}$ & $3.35^{+0.21}_{-0.12}$ & R \\
           LAMOST\,HVS\,1 &  8 &           $589^{+10}_{-13}$ &        $5.8^{+3.3}_{-2.5}$ & $8.53^{+0.27}_{-0.38}$ & D \\
                     B434 &  9 &           $586^{+17}_{-35}$ & $22^{+10}_{-\phantom{0}6}$ & $2.82^{+0.18}_{-0.07}$ & R \\
           LAMOST\,HVS\,2 & 10 &           $575^{+28}_{-47}$ &     $5.30^{+1.04}_{-0.25}$ &       $7.3 \pm \ldots$ & D \\
           PG\,1610$+$062 & 11 &           $561^{+15}_{-11}$ &        $6.2^{+0.5}_{-0.4}$ &    $4.4^{+0.1}_{-0.1}$ & D \\
                   HVS\,7 & 12 &           $561^{+16}_{-18}$ &             $22^{+4}_{-4}$ & $3.74^{+0.20}_{-0.11}$ & R \\
           LAMOST\,HVS\,4 & 13 &           $541^{+26}_{-47}$ &             $16^{+6}_{-4}$ &    $6.0^{+0.5}_{-0.5}$ & D \\
             J0136$+$2425 & 14 &           $500^{+50}_{-50}$ &       $14.2^{+1.8}_{-1.6}$ & $2.45^{+0.20}_{-0.20}$ & D \\
                     B733 & 15 &             $470^{+8}_{-9}$ &       $10.8^{+0.5}_{-0.5}$ & $2.72^{+0.06}_{-0.07}$ & D \\
                     B711 & 16 & $436^{+\phantom{0}6}_{-22}$ &             $22^{+5}_{-5}$ & $2.77^{+0.08}_{-0.06}$ & R \\
                  HVS\,17 & 17 & $436^{+12}_{-\phantom{0}9}$ &       $11.2^{+1.9}_{-1.6}$ & $3.43^{+0.10}_{-0.10}$ & D \\
                   HVS\,8 & 18 & $427^{+13}_{-\phantom{0}7}$ &       $11.6^{+3.3}_{-2.1}$ & $2.93^{+0.08}_{-0.07}$ & D \\
                     B485 & 19 &             $417^{+9}_{-6}$ &       $15.1^{+2.5}_{-1.7}$ & $5.05^{+0.45}_{-0.14}$ & D \\
              HIP\,114569 & 20 &           $414^{+18}_{-19}$ &     $5.44^{+0.23}_{-0.21}$ & $5.72^{+0.11}_{-0.10}$ & D \\
                 PHL\,346 & 21 &           $402^{+33}_{-30}$ &        $2.5^{+0.4}_{-0.4}$ & $9.05^{+0.28}_{-0.21}$ & D \\
               HIP\,60350 & 22 &           $388^{+15}_{-13}$ &     $6.28^{+0.08}_{-0.07}$ & $5.21^{+0.10}_{-0.10}$ & D \\
         EC\,04420$-$1908 & 23 &           $381^{+10}_{-12}$ &        $6.0^{+1.0}_{-0.7}$ & $6.20^{+0.24}_{-0.23}$ & D \\
                PHL\,2018 & 24 &           $374^{+19}_{-19}$ &     $5.07^{+0.22}_{-0.19}$ & $7.42^{+0.14}_{-0.53}$ & D \\
           LAMOST\,HVS\,3 & 25 &           $362^{+13}_{-10}$ &        $7.9^{+0.5}_{-0.6}$ & $3.78^{+0.12}_{-0.12}$ & D \\
                     B143 & 26 &           $360^{+50}_{-40}$ &        $8.1^{+1.2}_{-1.6}$ & $2.97^{+0.09}_{-0.08}$ & D \\
               HD\,271791 & 27 & $355^{+13}_{-\phantom{0}4}$ &       $11.8^{+1.4}_{-1.3}$ &   $10.9^{+0.4}_{-0.5}$ & D \\
         EC\,19596$-$5356 & 28 &           $355^{+10}_{-11}$ &       $14.8^{+0.9}_{-1.0}$ & $5.04^{+0.11}_{-0.08}$ & D \\
           BD\,$-$2\,3766 & 29 &           $349^{+18}_{-17}$ &     $8.69^{+0.17}_{-0.18}$ & $9.92^{+0.20}_{-0.20}$ & D \\
               HIP\,56322 & 30 &             $335^{+7}_{-6}$ &     $8.85^{+0.08}_{-0.08}$ & $9.78^{+0.20}_{-0.20}$ & D \\
\hline
\end{tabular}
\tablefoot{The given uncertainties are $1\sigma$ confidence intervals. The number in the second column is used to identify an object in Fig.~\ref{fig:vejection_versus_Galactocentric_radius}, Table~\ref{table:spectrophotometry}, and \ref{table:kinematics_AS}. The flag in the last column marks objects that potentially originate in the GC (``C''), are bona fide disk runaways (``D''), or stem from the outer rim of the disk (``R''). \tablefoottext{a}{Blue horizontal branch star.}}
\end{table}
\begin{figure*}
\includegraphics[width=1\textwidth]{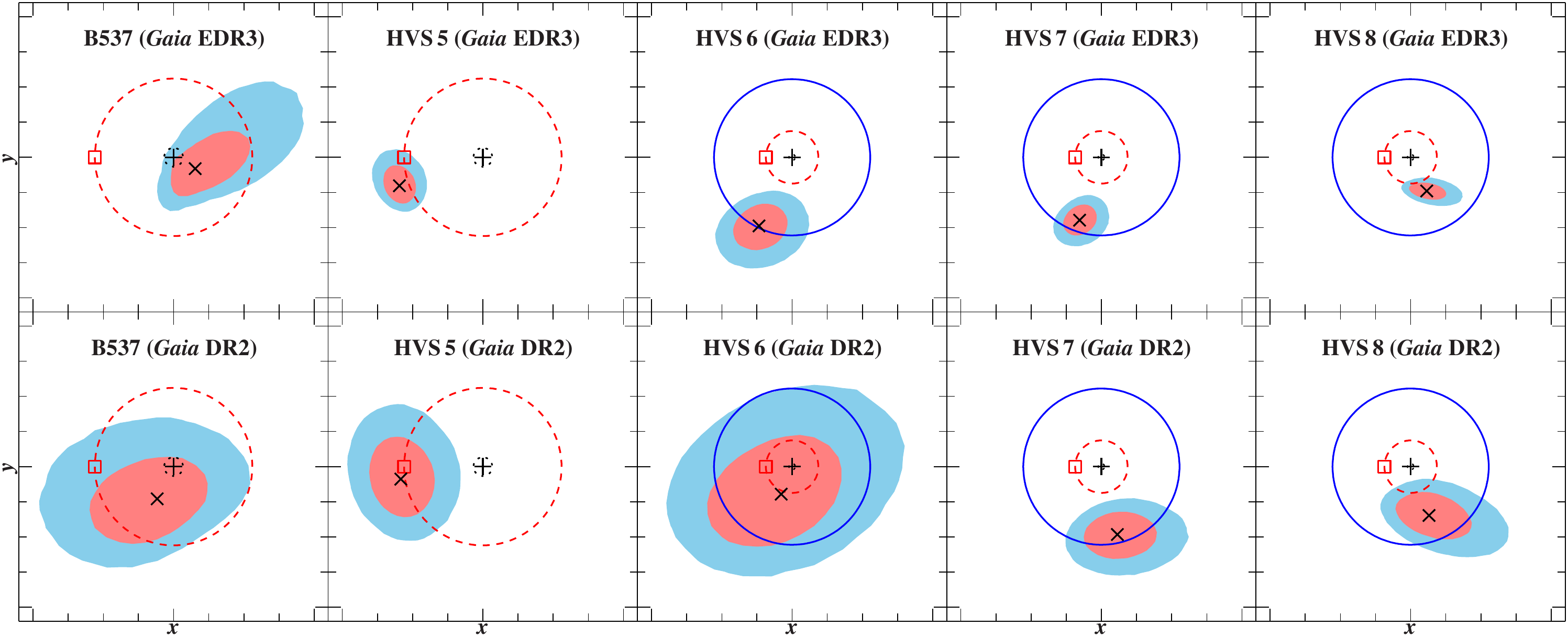}
\caption{\label{fig:plane_crossing_comparison_DR2_vs_EDR3}Inferred spatial origin within the Galactic plane for five selected objects (\textit{upper row}; see Fig.~\ref{fig:plane_crossing} for all program stars). The most likely plane-crossing point is marked by a black cross, while the shaded areas visualize the corresponding $1\sigma$ (light red) and $2\sigma$ (light blue) contours. Circles centered at the GC (black plus sign) with radii of 1\,kpc (dotted black line), 8.3\,kpc (solar circle; dashed red line), and 25\,kpc (solid blue line) are shown for reference. The galactocentric coordinate system is Cartesian and right-handed, with the Sun (red square) on the negative $x$-axis and the $z$-axis pointing to the Galactic north pole. The panels in the \textit{lower row} show the same, except that proper motions from {\it Gaia} DR2 are used as input, demonstrating the improved precision of EDR3.}
\end{figure*}
\section{\label{sect:analysis}Quantitative spectrophotometric analysis}
Although {\it Gaia} EDR3 provides astrometric data of unprecedented precision and accuracy, distances estimated from the measured parallaxes are still considerably less precise than those obtained from spectrophotometry, which have typical uncertainties smaller than 10\%, mainly because our targets are found at distances beyond the current capabilities of {\it Gaia}. Consequently, we used spectrophotometric distance estimates as input for our kinematic investigation. For half of the sample, that is, for the 14 stars from the MMT-HVS survey and for PG\,1610+062, we used our previously published results (see \citeads{2019A&A...628L...5I}; \citeads{2020A&A...637A..53K}). Seeking homogeneity, we also carried out quantitative spectral analyses of the remaining targets using the same techniques as applied for the others. This was possible for all stars except S5-HVS\,1 and LAMOST HVS\,2, for which we were not able to obtain spectra and thus had to rely on literature values. We also used the latter for J0136$+$2425 because the atmospheric parameters inferred by us do not deviate from the published ones by \citetads{2009A&A...507L..37T}.

Our analysis strategy and the adopted model atmospheres are extensively outlined in previous papers (\citeads{2014A&A...565A..63I}; \citeads{2018OAst...27...35H}; \citeads{2018A&A...615L...5I}; \citeads{2019A&A...628L...5I}; \citeads{2020A&A...637A..53K}). We therefore briefly summarize the key details here. Atmospheric parameters such as effective temperature $T_{\mathrm{eff}}$, surface gravity $\log(g)$, projected rotational velocity $\varv\sin(i)$, and radial velocity $\varv_\mathrm{rad}$ were determined by fitting the most recent model spectra of appropriate chemical composition to observed spectra. The resulting parameters were then used to compute a synthetic spectral energy distribution that was fit to measured magnitudes, which usually cover the ultraviolet, optical, and infrared regime, in order to derive the angular diameter $\Theta$ of the star and the interstellar extinction. The latter is typically very small for our program stars owing to their position above or below the Galactic disk. Stellar masses $M$ and ages $\tau_\mathrm{evol}$ were inferred by comparing the position of a star in the Kiel diagram to evolutionary tracks by the Geneva group (\citeads{2012A&A...537A.146E}; \citeads{2013A&A...553A..24G}). The spectrophotometric distance $d$ finally followed from solving the definition of the surface gravity, $g = GMR^{-2}$ , with $G$ being the gravitational constant, for the stellar radius $R$ and inserting the latter in the definition of the angular diameter, $\Theta = 2R/d$. The resulting atmospheric and stellar parameters as well as information about the available optical spectra are summarized in Table~\ref{table:spectrophotometry}.
\section{\label{sect:kinematics} \textit{Gaia} EDR3 update on kinematics}
The spectrophotometric distances and radial velocities were combined with proper motions from {\it Gaia} EDR3 to trace back the trajectories of the program stars to the Galactic plane in order to derive their ejection velocities with respect to the rotating Galactic disk and to identify their spatial origins within the disk.

To this end, we used the Milky Way mass model~I of \citetads{2013A&A...549A.137I}, which is a three-component Galactic potential with standard analytical forms for the central spherical bulge, an axisymmetric disk, and a spherical dark matter halo. The model reproduces a variety of observational constraints such as the Galactic rotation curve, and according to \citetads{2018A&A...620A..48I}, is still valid with Milky Way mass measurements based on the motions of globular clusters, satellite galaxies, and extreme velocity halo stars as derived from {\it Gaia} DR2 astrometry. The corresponding equations of motions are numerically solved with a fourth-order Runge-Kutta-Fehlberg method with adaptive stepsize control that ensures an absolute numerical accuracy in each integration step of less than $10^{-10}$ for positions in kpc and velocities in km\,s$^{-1}$. Error propagation for the four input parameters (spectrophotometric distance, radial velocity, and proper motion in right ascension and declination) is handled with a Monte Carlo simulation with one million runs, using Gaussian distributions that allow for asymmetric error bars and the correlation between the two proper motion components. Uncertainties in the distance of the Sun to the GC and in its relative motion to the local standard of rest \citepads{2010MNRAS.403.1829S} are treated analogously. To account for the nonzero height of the Galactic disk, the backward integration of the Monte Carlo orbits is not stopped exactly at the Galactic mid-plane, but at vertical offsets randomly drawn from a Gaussian probability distribution with a mean value of zero and a standard deviation of 0.1\,kpc (motivated, e.g., by \citeads{2014MNRAS.444..290B}; \citeads{2019ApJ...871..208L}). Because stars in the outer disk have been observed at much larger vertical distances of up to several kpc (see, e.g., \citeads{2014Natur.509..342F}; \citeads{2015MNRAS.452..676P}; \citeads{2018Natur.555..334B}), we used a standard deviation of 5\,kpc for targets originating from galactocentric radii larger than 16\,kpc. Owing to the high space velocities of our targets, the derived properties at plane intersection are only marginally affected by considering a nonzero height for the disk, however.

\begin{figure}
\includegraphics[width=1\linewidth]{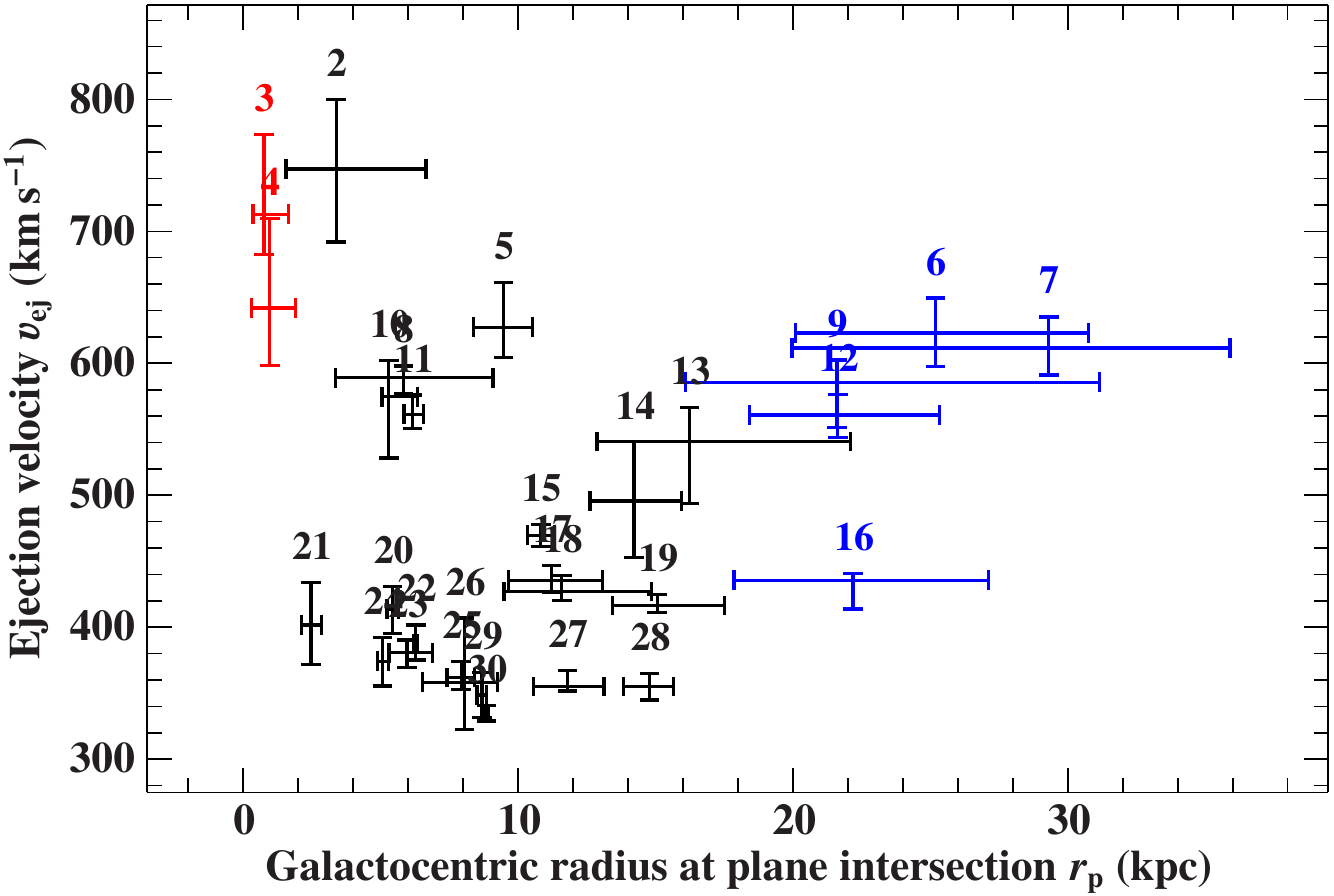}
\caption{\label{fig:vejection_versus_Galactocentric_radius}Ejection velocity (relative to the rotating Galactic disk) vs.\ galactocentric radius at plane intersection for all program stars except S5-HVS\,1, which is beyond the ordinate scale. Error bars are $1\sigma$ confidence intervals. The identification numbers above the data points correspond to those in Table~\ref{table:vejection}. The colors mark objects that could stem from the GC (red) or from the outer disk (blue), see Fig.~\ref{fig:plane_crossing}.}
\end{figure}

To demonstrate the immense gain in precision from DR2 to EDR3, we show the inferred plane-crossing areas for five selected targets for both cases in Fig.~\ref{fig:plane_crossing_comparison_DR2_vs_EDR3}. While all resulting kinematic parameters are summarized in Table~\ref{table:kinematics_AS}, we focus in the following on the ejection velocities $\varv_{\mathrm{ej}}$ and galactocentric radii at plane intersection $r_\mathrm{p}$, which are also given in Table~\ref{table:vejection} or are visualized in Fig.~\ref{fig:vejection_versus_Galactocentric_radius}. Three subgroups may be identified in our sample: stars originating in the GC, bona fide disk runaways, and stars that were presumably ejected from the outer rim of the Galactic disk.
\subsection{Possible origin in the Galactic center}
Only three program stars have galactocentric radii at plane intersection that are smaller than 1\,kpc and may thus stem from the GC: S5-HVS\,1, B576, and B598. These three are also among the top four with the highest ejection velocities, with S5-HVS\,1 clearly standing out with $\varv_\mathrm{ej} = 1810\pm60$\,km\,s$^{-1}$ . It is also the only object that is gravitationally unbound to the Milky Way. Within the error bars, our result for this object agrees with that of \citetads{2020MNRAS.491.2465K}, which is expected because we used the same input for the kinematic investigation, except for updated proper motions, and the latter have not changed significantly between DR2 and EDR3. The conclusions of \citetads{2020MNRAS.491.2465K} are therefore still valid and S5-HVS\,1 can be considered as evidence for the Hills mechanism. Changes in proper motions for the other two objects in this subgroup, B576 and B598, are also negligible, which is why the results reported by \citetads{2020A&A...637A..53K} are basically confirmed here. However, we note that the nature of these two stars is not yet clear\footnote{ \citetads{2020A&A...637A..53K} argued that B576 is probably a blue horizontal branch star, but an MS nature could not be ruled out. B598 was speculated to be the progenitor of an extremely low mass white dwarf, but an MS nature was again a viable option. Although still not very precise, the parallaxes from EDR3 keep favoring the blue-horizontal-branch scenario for B576, but now at least tend to prefer an MS scenario for B598. If the latter were indeed an MS star, it had to be rejuvenated in the course of its evolution to reconcile its inferred stellar age ($5\pm1$\,Myr) with its flight time ($50\pm5$\,Myr), and it may therefore be a blue straggler.}.

Interestingly, the ejection velocities of all program stars are lower than half of the ejection velocity of S5-HVS\,1. \citetads{2020A&A...637A..53K} argued that HVS\,22 (\object{{\it Gaia} EDR3 3897063727354575488}), which is not part of the current sample because its origin is still not constrained at all, could be similarly extreme, but our reanalysis based on EDR3 proper motions led to a substantially lower ejection velocity of $720^{+110}_{-130}$\,km\,s$^{-1}$ instead of $1510^{+680}_{-550}$\,km\,s$^{-1}$, demonstrating that S5-HVS\,1 is a truly exceptional case.
\subsection{Origin in the Galactic disk}
This group, which represents the vast majority of the program stars, contains objects that have past trajectories that unambiguously indicate an origin in the Galactic disk. A possible but unlikely exception is B537, which is also included in this group although its $2\sigma$ plane-crossing contour covers the GC. For this particular target, the change in proper motions from DR2 to EDR3 significantly affects the inferred place of origin, see Fig.~\ref{fig:plane_crossing_comparison_DR2_vs_EDR3}, as well as the deduced ejection velocity, $750\pm60$\,km\,s$^{-1}$ instead of $460^{+210}_{-\phantom{0}90}$\,km\,s$^{-1}$, making it the target with the second highest ejection velocity in the entire sample. Although slightly less extreme in terms of ejection velocity ($627^{+34}_{-24}$\,km\,s$^{-1}$), HVS\,5 is certainly one of the most interesting targets because it is, on the one hand, clearly gravitationally unbound to the Milky Way, and on the other hand, its place of origin is now so precisely constrained to lie in the current solar neighborhood that a possible origin in the GC, as statistically favored by \citetads{2018ApJ...866...39B}, can be definitely ruled out. LAMOST HVS\,1 and LAMOST HVS\,4 are two further examples for such so-called hyper-runaway stars. For the former, we note that our revised analysis solves the difference between stellar age and flight time as reported by \citetads{2019ApJ...873..116H}. Interestingly, the object for which the hyper-runaway class was originally coined, HD\,271791, no longer qualifies as member because it is clearly bound to the Milky Way. The same applies to HIP\,60350. Whether LAMOST HVS\,2, J0136$+$2425, and HVS\,8 are hyper-runaway stars depends on the choice of the applied Galactic mass model. The higher precision of {\it Gaia} EDR3 also allows us to strengthen our previous argument that the well-studied MS disk-runaway star PG\,1610$+$062 is a serious challenge for classical ejection scenarios \citepads{2019A&A...628L...5I}.
\subsection{Possible origin in the outer rim of the Galactic disk}
It is intriguing that a group of five stars (HVS\,6, B481, B434, HVS\,7, and B711) appears to separate itself from the others with respect to the galactocentric radius at plane intersection $r_\mathrm{p}$, see Fig.~\ref{fig:vejection_versus_Galactocentric_radius}. The distribution of the most likely value for $r_\mathrm{p}$ is more or less continuous up to $\sim$15\,kpc, a number that fairly well coincides with the observed dimension of the spiral arms \citepads{2014A&A...569A.125H}, then there is a gap of about 6\,kpc before another group of runaway stars appears. Moreover, it is interesting to note that the ejection velocities of four of the five outer-rim stars are higher than those of most of the objects that are classified as disk-runaway stars in Table~\ref{table:vejection}. Although based on low-number statistics and data points with relatively large error bars, this observation might indicate a distinct population of runaway stars with regard to spatial origin and ejection mechanism. Overdensities in MS star counts in form of concentric rings around the GC with radii of at least 25\,kpc have been observed (see \citeads{2015ApJ...801..105X} and references therein), which could be the origin of the five targets discussed in this paragraph. The same scenario, that is, the accretion of small satellite galaxies, has been suggested as possible explanation for the presence of the rings \citepads{2015ApJ...801..105X} and high-velocity runaway stars \citepads{2009ApJ...691L..63A}. Our results may not only indicate that this mechanism is in fact operating, but is also able to produce HVSs because HVS\,6 is unbound to the Milky Way while B481 and HVS\,7 might be.
\section{Summary and conclusion}
Based on partly new spectrophotometric distances and the unprecedented precision of proper motions from {\it Gaia} EDR3, we have revisited the kinematics of 30 of the most extreme blue runaway candidates known to date in order to derive their ejection velocity from and their spatial origin within the Galactic disk. In addition to S5-HVS\,1, which can still be considered as evidence for the ejection of stars from the GC through the Hills mechanism, there are only two other targets, B576 and B598, for which an origin in the GC is likely, provided that their stellar nature is confirmed by future analyses of currently unavailable high-quality spectra. For a fourth object, B537, the place of origin is still poorly constrained, therefore we cannot rule out a GC origin.

Despite these four stars, {\it Gaia} EDR3 corroborates the trend that most of the former Hills candidates turn out to be extreme runaway stars from the Galactic disk when their kinematic properties are sufficiently known. This particularly applies to the gravitationally unbound stars HVS\,5 and HVS\,6 here, for which the GC no longer is a valid birth place. Moreover, all program stars have such high ejection velocities that they considerably differ from the classical scenarios for an ejection from the Galactic disk, that is, the BSS and the DES. This strengthens the idea that mechanisms that involve dynamical encounters with very massive stars or intermediate-mass black holes are responsible for the majority of the most extreme runaway stars. The latter option is particularly interesting because no intermediate-mass black hole is currently known in the Galactic disk \citepads{2020ARA&A..58..257G}. The inferred spatial origins of extreme disk-runaway stars may give useful indications of where to search for them.

Five stars in the sample have galactocentric radii at disk intersection that are larger than 21\,kpc, which implies that their birth places are probably not related to the spiral arms but to the outer Galactic rings. The ejection velocities of these five stars are also mostly higher than those of the inner disk runaways. Because high-velocity runaway stars and Galactic rings were suggested to result from the accretion of small satellite galaxies, this finding may be taken as another indication for the variety of mechanisms that produce runaway stars.
\begin{acknowledgements}
We thank Adrian M.~Price-Whelan for a very constructive referee report. A.I.\ and U.H.\ acknowledge funding by the Deutsche For\-schungs\-gemeinschaft (DFG) through grants IR190/1-1, HE1356/70-1, and HE1356/71-1. R.R.\ has received funding from the postdoctoral fellowship program Beatriu de Pin\'os, funded by the Secretary of Universities and Research (Government of Catalonia) and by the Horizon 2020 program of research and innovation of the European Union under the Maria Sk\l{}odowska-Curie grant agreement No 801370. We thank John E.\ Davis for the development of the {\sc slxfig} module used to prepare the figures in this paper. This work has made use of data from the European Space Agency (ESA) mission {\it Gaia} (\url{https://www.cosmos.esa.int/gaia}), processed by the {\it Gaia} Data Processing and Analysis Consortium (DPAC, \url{https://www.cosmos.esa.int/web/gaia/dpac/consortium}). Funding for the DPAC has been provided by national institutions, in particular the institutions participating in the {\it Gaia} Multilateral Agreement. Based on observations made with ESO Telescopes at the La Silla Paranal Observatory under programme IDs 067.D-0010(A), 088.A-9003(A), 088.D-0064(A), 091.D-0061(A), 093.D-0302(A), 282.D-5065(A), and 383.D-0909(A). Based on observations made with the Nordic Optical Telescope, operated by the Nordic Optical Telescope Scientific Association at the Observatorio del Roque de los Muchachos, La Palma, Spain, of the Instituto de Astrofisica de Canarias. Guoshoujing Telescope (the Large Sky Area Multi-Object Fiber Spectroscopic Telescope LAMOST) is a National Major Scientific Project built by the Chinese Academy of Sciences. Funding for the project has been provided by the National Development and Reform Commission. LAMOST is operated and managed by the National Astronomical Observatories, Chinese Academy of Sciences.
\end{acknowledgements}
\bibliographystyle{aa}

\noindent\flushcolsend
\begin{appendix}
\renewcommand{\thefigure}{A.\arabic{figure}}
\renewcommand{\thetable}{A.\arabic{table}}
\renewcommand{\theequation}{A.\arabic{equation}}
\begin{figure*}
\includegraphics[width=1\textwidth]{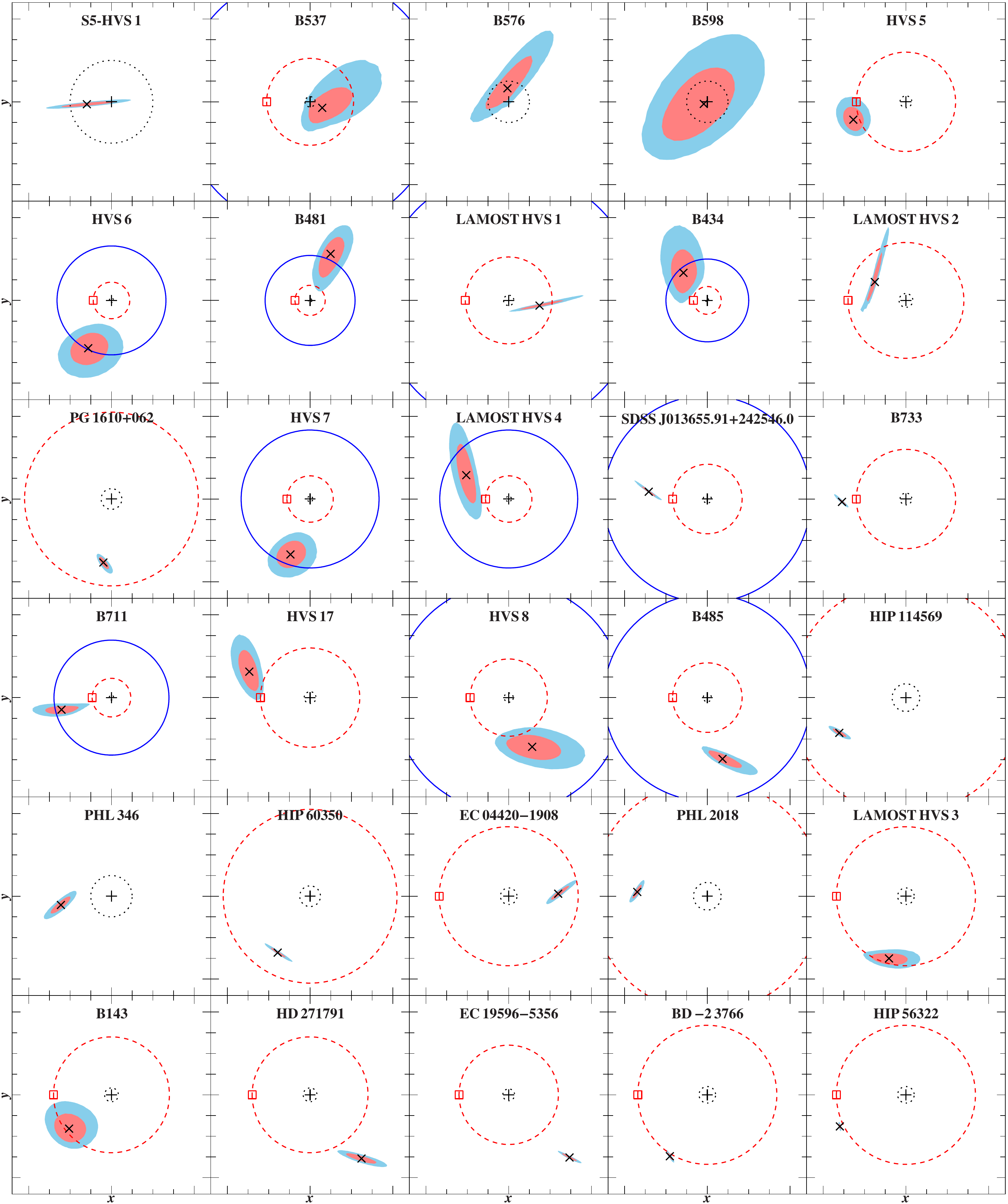}
\caption{\label{fig:plane_crossing}Inferred spatial origin within the Galactic plane for all program stars. The stars are arranged in the same order as in Table~\ref{table:vejection}, that is, with decreasing ejection velocity starting from the top left panel. As in Fig.~\ref{fig:plane_crossing_comparison_DR2_vs_EDR3}, the most likely plane-crossing point is marked by a black cross, and the shaded areas visualize the corresponding $1\sigma$ (light red) and $2\sigma$ (light blue) contours. Note the different scales: Circles centered at the GC (black plus sign) with radii of 1\,kpc (dotted black line), 8.3\,kpc (solar circle; dashed red line), and 25\,kpc (solid blue line) are shown for reference, when appropriate. The current position of the Sun (red square) is also marked.}
\end{figure*}
{
\clearpage
\onecolumn
\begin{landscape}
\begin{table*}
\small
\centering
\renewcommand{\arraystretch}{1.33}
\caption{\label{table:spectrophotometry}Atmospheric and stellar parameters of the program stars.}
\begin{tabular}{rl|ccccc|ccccc|c}
\hline\hline
& & \multicolumn{5}{c|}{Atmospheric parameters} & \multicolumn{5}{c|}{Stellar parameters} & \\
\hline
\# & {\it Gaia} EDR3 identifier & $T_{\mathrm{eff}}$ & $\log(g)$ & $\varv\sin(i)$ & $\varv_\mathrm{rad}$ & Spectrograph\,\tablefootmark{(a)} & $M$ & $R$ & $\log(L/L_\odot)$ & $\tau_\mathrm{evol}$ & $d$ & Ref.\,\tablefootmark{(b)} \\
\cline{5-6}
& & (K) & (cgs) & \multicolumn{2}{c}{(km\,s$^{-1}$)} & & ($M_\odot$) & ($R_\odot$) & & (Myr) & (kpc) \\
\hline
 1 & \object{{\it\,Gaia}\,EDR3\,6513109241989477504} & $\phantom{0}9\,630^{+110}_{-110}$ & $4.23^{+0.03}_{-0.03}$ &                \ldots &   $1017.0^{+2.7}_{-2.7}$ & A\,(1300, 10\,000) & $2.35^{+0.06}_{-0.06}$ & $1.95^{+0.08}_{-0.08}$ & $1.47^{+0.04}_{-0.04}$ &            $52^{+41}_{-28}$ &  $8.63^{+0.31}_{-0.30}$ & KO \\
 2 & \object{{\it\,Gaia}\,EDR3\,2799690344752115584} &           $11\,760^{+170}_{-190}$ & $3.75^{+0.08}_{-0.08}$ &     $181^{+40}_{-30}$ &    $150.9^{+9.5}_{-9.6}$ &      M\,(1.2\,\AA) & $3.73^{+0.26}_{-0.17}$ &    $4.3^{+0.6}_{-0.5}$ & $2.50^{+0.12}_{-0.10}$ &           $190^{+17}_{-34}$ &          $41^{+6}_{-5}$ & KR \\
 3 & \object{{\it\,Gaia}\,EDR3\,1482227680963943808} &           $11\,400^{+130}_{-130}$ & $3.70^{+0.05}_{-0.05}$ &      $47^{+15}_{-15}$ &    $216.1^{+2.0}_{-1.8}$ &      M\,(1.2\,\AA) &         $0.5\pm$\ldots & $1.65^{+0.10}_{-0.10}$ & $1.62^{+0.05}_{-0.06}$ &                      \ldots &    $18.7^{+1.6}_{-1.5}$ & KR \\
 4 & \object{{\it\,Gaia}\,EDR3\,1225026618163072512} &           $10\,730^{+130}_{-130}$ & $4.52^{+0.05}_{-0.05}$ &     $192^{+19}_{-16}$ &    $282.5^{+4.7}_{-5.1}$ &      M\,(1.2\,\AA) & $2.40^{+0.05}_{-0.05}$ &    $1.6^{+0.1}_{-0.1}$ & $1.53^{+0.03}_{-0.04}$ &               $5^{+1}_{-1}$ &          $23^{+2}_{-2}$ & KR \\
 5 & \object{{\it\,Gaia}\,EDR3\,1069326945513133952} &           $12\,550^{+160}_{-150}$ & $4.09^{+0.06}_{-0.06}$ &     $123^{+19}_{-19}$ &    $541.5^{+6.4}_{-6.1}$ &      M\,(1.2\,\AA) & $3.40^{+0.10}_{-0.10}$ &    $2.8^{+0.3}_{-0.3}$ & $2.23^{+0.08}_{-0.08}$ &           $149^{+20}_{-26}$ &          $37^{+4}_{-4}$ & KR \\
 6 & \object{{\it\,Gaia}\,EDR3\,3867267443277880320} &           $12\,390^{+160}_{-160}$ & $4.30^{+0.06}_{-0.06}$ &      $79^{+25}_{-25}$ &    $606.3^{+6.5}_{-6.2}$ &      M\,(1.2\,\AA) & $3.04^{+0.09}_{-0.09}$ &    $2.0^{+0.2}_{-0.2}$ & $1.95^{+0.08}_{-0.08}$ &            $62^{+41}_{-47}$ &          $52^{+6}_{-5}$ & KR \\
 7 & \object{{\it\,Gaia}\,EDR3\,2660715403600483584} &           $10\,300^{+170}_{-200}$ & $3.61^{+0.08}_{-0.08}$ &     $190^{+25}_{-31}$ &    $133.1^{+9.2}_{-8.5}$ &      M\,(1.2\,\AA) & $3.35^{+0.21}_{-0.12}$ &    $4.8^{+0.6}_{-0.5}$ & $2.36^{+0.10}_{-0.09}$ &           $279^{+31}_{-54}$ &          $46^{+5}_{-6}$ & KR \\
 8 &  \object{{\it\,Gaia}\,EDR3\,590511484409775360} &           $19\,180^{+380}_{-380}$ & $3.58^{+0.08}_{-0.08}$ & $161.5^{+1.8}_{-1.5}$ &    $612.2^{+5.7}_{-5.7}$ &          L\,(1800) & $8.53^{+0.27}_{-0.38}$ &    $7.8^{+0.8}_{-0.8}$ & $3.88^{+0.09}_{-0.10}$ &        $35.1^{+2.8}_{-2.7}$ &    $15.7^{+1.5}_{-1.5}$ &  C \\
 9 & \object{{\it\,Gaia}\,EDR3\,3814622895259904256} &           $10\,140^{+120}_{-120}$ & $3.84^{+0.05}_{-0.05}$ &      $89^{+12}_{-16}$ &    $441.3^{+2.9}_{-2.9}$ &      M\,(1.2\,\AA) & $2.82^{+0.18}_{-0.07}$ &    $3.3^{+0.3}_{-0.3}$ & $2.02^{+0.07}_{-0.06}$ &           $388^{+16}_{-96}$ &          $41^{+4}_{-3}$ & KR \\
10 & \object{{\it\,Gaia}\,EDR3\,1330715287893559936} &                $20\,600\pm$\ldots &        $4.13\pm$\ldots &                \ldots & $341.10^{+7.79}_{-7.79}$ &          L\,(1800) &         $7.3\pm$\ldots &        $3.84\pm$\ldots &        $3.38\pm$\ldots &               $16\pm$\ldots & $22.24^{+4.57}_{-4.57}$ &  H \\
11 & \object{{\it\,Gaia}\,EDR3\,4450123955938796160} &           $14\,800^{+120}_{-120}$ & $4.05^{+0.04}_{-0.04}$ &  $15.5^{+0.6}_{-0.7}$ &        $157.4^{+3}_{-3}$ &          E\,(8000) &    $4.4^{+0.1}_{-0.1}$ &    $3.3^{+0.2}_{-0.2}$ & $2.66^{+0.05}_{-0.04}$ &  $83^{+\phantom{0}9}_{-10}$ &    $17.3^{+1.2}_{-1.0}$ &  I \\
12 & \object{{\it\,Gaia}\,EDR3\,3799146650623432704} &           $12\,950^{+140}_{-140}$ & $3.96^{+0.05}_{-0.05}$ &      $52^{+16}_{-15}$ &    $521.8^{+2.6}_{-2.5}$ &      M\,(1.2\,\AA) & $3.74^{+0.20}_{-0.11}$ &    $3.3^{+0.3}_{-0.3}$ & $2.45^{+0.09}_{-0.08}$ & $152^{+\phantom{0}9}_{-16}$ &          $48^{+5}_{-4}$ & KR \\
13 & \object{{\it\,Gaia}\,EDR3\,1928660566125735680} &           $13\,520^{+270}_{-270}$ & $3.25^{+0.08}_{-0.08}$ &     $148^{+19}_{-20}$ &    $332.9^{+4.5}_{-5.0}$ &          L\,(1800) &    $6.0^{+0.5}_{-0.5}$ &    $9.6^{+1.0}_{-1.0}$ & $3.44^{+0.10}_{-0.10}$ &            $83^{+14}_{-14}$ &          $69^{+8}_{-7}$ &  C \\
14 &  \object{{\it\,Gaia}\,EDR3\,291821209329550464} & $\phantom{0}9\,100^{+250}_{-250}$ & $3.90^{+0.15}_{-0.15}$ &        $250\pm$\ldots &    $324.3^{+5.9}_{-5.9}$ &          S\,(2000) & $2.45^{+0.20}_{-0.20}$ &    $2.9^{+0.5}_{-0.6}$ & $1.72^{+0.17}_{-0.17}$ &              $245\pm$\ldots &    $10.9^{+2.0}_{-2.0}$ &  T \\
15 & \object{{\it\,Gaia}\,EDR3\,1283080527168129536} &           $10\,280^{+120}_{-120}$ & $4.04^{+0.05}_{-0.05}$ &     $278^{+12}_{-10}$ &    $351.4^{+3.3}_{-2.7}$ &      M\,(1.2\,\AA) & $2.72^{+0.06}_{-0.07}$ &    $2.6^{+0.2}_{-0.2}$ & $1.83^{+0.06}_{-0.06}$ &           $273^{+97}_{-36}$ &          $12^{+1}_{-1}$ & KR \\
16 & \object{{\it\,Gaia}\,EDR3\,1226051053762724480} &           $10\,410^{+120}_{-120}$ & $3.98^{+0.05}_{-0.05}$ &      $17^{+20}_{-17}$ &    $267.6^{+2.2}_{-2.5}$ &      M\,(1.2\,\AA) & $2.77^{+0.08}_{-0.06}$ &    $2.8^{+0.2}_{-0.2}$ & $1.92^{+0.06}_{-0.06}$ &           $315^{+62}_{-21}$ &          $23^{+2}_{-2}$ & KR \\
17 & \object{{\it\,Gaia}\,EDR3\,1407293627068696192} &           $12\,620^{+140}_{-140}$ & $4.09^{+0.05}_{-0.05}$ &     $129^{+14}_{-16}$ &    $255.5^{+3.4}_{-3.8}$ &      M\,(1.2\,\AA) & $3.43^{+0.10}_{-0.10}$ &    $2.8^{+0.2}_{-0.2}$ & $2.24^{+0.07}_{-0.07}$ &           $143^{+20}_{-20}$ &          $35^{+3}_{-3}$ & KR \\
18 &  \object{{\it\,Gaia}\,EDR3\,633599760258827776} &           $10\,880^{+120}_{-120}$ & $4.06^{+0.05}_{-0.05}$ &     $276^{+14}_{-11}$ &    $498.9^{+3.6}_{-3.4}$ &      M\,(1.2\,\AA) & $2.93^{+0.08}_{-0.07}$ &    $2.7^{+0.2}_{-0.2}$ & $1.95^{+0.07}_{-0.07}$ &           $208^{+54}_{-33}$ &          $36^{+3}_{-3}$ & KR \\
19 &  \object{{\it\,Gaia}\,EDR3\,742363828436051456} &           $15\,710^{+180}_{-170}$ & $3.92^{+0.05}_{-0.05}$ &      $81^{+12}_{-14}$ &    $422.7^{+2.1}_{-2.7}$ &      M\,(1.2\,\AA) & $5.05^{+0.45}_{-0.14}$ &    $4.1^{+0.5}_{-0.3}$ & $2.96^{+0.09}_{-0.07}$ &  $79^{+\phantom{0}4}_{-18}$ &          $33^{+4}_{-3}$ & KR \\
20 & \object{{\it\,Gaia}\,EDR3\,2386639629044028160} &           $17\,350^{+170}_{-170}$ & $4.04^{+0.04}_{-0.04}$ &  $83.0^{+0.8}_{-0.8}$ &     $99.7^{+1.5}_{-1.5}$ &      FE\,(48\,000) & $5.72^{+0.11}_{-0.10}$ & $3.78^{+0.18}_{-0.18}$ & $3.07^{+0.05}_{-0.05}$ &        $52.8^{+2.6}_{-3.8}$ &  $1.64^{+0.08}_{-0.08}$ &  C \\
21 & \object{{\it\,Gaia}\,EDR3\,2402031280004432512} &           $21\,570^{+220}_{-220}$ & $3.83^{+0.04}_{-0.04}$ &  $27.3^{+0.3}_{-0.3}$ &     $59.9^{+1.2}_{-1.2}$ &       U\,(35\,690) & $9.05^{+0.28}_{-0.21}$ & $6.06^{+0.29}_{-0.30}$ & $3.86^{+0.05}_{-0.05}$ &        $23.2^{+1.0}_{-0.7}$ &     $6.5^{+0.4}_{-0.4}$ &  C \\
22 & \object{{\it\,Gaia}\,EDR3\,1533367925276710272} &           $16\,520^{+170}_{-170}$ & $4.08^{+0.04}_{-0.04}$ & $135.1^{+0.2}_{-0.4}$ &    $263.6^{+0.4}_{-0.5}$ &      FI\,(45\,000) & $5.21^{+0.10}_{-0.10}$ & $3.44^{+0.17}_{-0.17}$ & $2.90^{+0.05}_{-0.05}$ &              $55^{+5}_{-6}$ &  $3.35^{+0.16}_{-0.16}$ &  C \\
23 & \object{{\it\,Gaia}\,EDR3\,2977856789466365056} &           $12\,990^{+130}_{-130}$ & $3.16^{+0.04}_{-0.04}$ & $228.2^{+2.3}_{-2.3}$ &    $211.1^{+2.5}_{-2.5}$ &      FE\,(48\,000) & $6.20^{+0.24}_{-0.23}$ &   $10.8^{+0.6}_{-0.6}$ & $3.48^{+0.05}_{-0.05}$ &              $78^{+8}_{-8}$ &    $17.0^{+0.9}_{-0.8}$ &  C \\
24 & \object{{\it\,Gaia}\,EDR3\,6624328240456787840} &           $18\,690^{+190}_{-190}$ & $3.78^{+0.04}_{-0.04}$ & $240.9^{+2.4}_{-2.4}$ &    $145.3^{+3.7}_{-3.7}$ &      FE\,(48\,000) & $7.42^{+0.14}_{-0.53}$ & $5.77^{+0.29}_{-0.31}$ & $3.56^{+0.05}_{-0.05}$ &        $40.1^{+7.9}_{-1.2}$ &     $6.7^{+0.4}_{-0.4}$ &  C \\
25 &   \object{{\it\,Gaia}\,EDR3\,56282900715073664} &           $12\,980^{+260}_{-260}$ & $4.02^{+0.08}_{-0.08}$ &     $228^{+19}_{-23}$ &    $360.9^{+5.1}_{-4.2}$ &          L\,(1800) & $3.78^{+0.12}_{-0.12}$ & $3.13^{+0.30}_{-0.29}$ & $2.40^{+0.09}_{-0.09}$ &           $138^{+13}_{-19}$ &    $22.1^{+2.1}_{-2.1}$ &  C \\
26 & \object{{\it\,Gaia}\,EDR3\,1034833616482496256} &           $10\,910^{+130}_{-120}$ & $4.03^{+0.05}_{-0.05}$ &     $269^{+22}_{-19}$ &    $217.6^{+5.4}_{-4.9}$ &      M\,(1.2\,\AA) & $2.97^{+0.09}_{-0.08}$ &    $2.8^{+0.3}_{-0.2}$ & $1.99^{+0.07}_{-0.07}$ &           $215^{+25}_{-32}$ &          $29^{+3}_{-3}$ & KR \\
27 & \object{{\it\,Gaia}\,EDR3\,5284151216932205312} &           $18\,630^{+190}_{-190}$ & $3.17^{+0.04}_{-0.04}$ & $127.0^{+0.1}_{-0.1}$ &    $442.5^{+0.4}_{-0.4}$ &       U\,(31\,585) &   $10.9^{+0.4}_{-0.5}$ &   $14.2^{+0.8}_{-0.7}$ & $4.34^{+0.05}_{-0.05}$ &        $23.8^{+1.8}_{-1.8}$ &    $19.6^{+1.0}_{-1.0}$ &  C \\
28 & \object{{\it\,Gaia}\,EDR3\,6473211813308625024} &           $15\,900^{+160}_{-160}$ & $4.05^{+0.04}_{-0.04}$ & $236.5^{+2.4}_{-2.4}$ &   $-202.6^{+0.3}_{-0.3}$ &       U\,(41\,640) & $5.04^{+0.11}_{-0.08}$ & $3.50^{+0.17}_{-0.16}$ & $2.85^{+0.05}_{-0.05}$ &              $60^{+5}_{-6}$ &    $11.5^{+0.6}_{-0.6}$ &  C \\
29 & \object{{\it\,Gaia}\,EDR3\,3657546118655166080} &           $23\,620^{+240}_{-240}$ & $3.99^{+0.04}_{-0.04}$ & $191.5^{+1.9}_{-1.9}$ &     $25.2^{+1.0}_{-1.0}$ &       U\,(28\,445) & $9.92^{+0.20}_{-0.20}$ & $5.27^{+0.25}_{-0.25}$ & $3.89^{+0.05}_{-0.05}$ &        $16.7^{+0.9}_{-1.2}$ &  $3.46^{+0.17}_{-0.17}$ &  C \\
30 & \object{{\it\,Gaia}\,EDR3\,3813323860926633344} &           $24\,020^{+240}_{-240}$ & $4.10^{+0.04}_{-0.04}$ & $172.9^{+1.7}_{-1.7}$ &    $260.3^{+1.3}_{-1.3}$ &       U\,(28\,445) & $9.78^{+0.20}_{-0.20}$ & $4.61^{+0.23}_{-0.22}$ & $3.81^{+0.05}_{-0.05}$ &        $12.5^{+1.7}_{-2.0}$ &  $3.10^{+0.15}_{-0.15}$ &  C \\
\hline
\end{tabular}
\tablefoot{The identification numbers in the first column correspond to those in Table~\ref{table:vejection}. The given uncertainties are $1\sigma$ confidence intervals. \tablefoottext{a}{The atmospheric parameters are based on spectra taken with instruments abbreviated as follows: A (AAOmega, \citeads{2006SPIE.6269E..0GS}), E (ESI, \citeads{2002PASP..114..851S}), FE (FEROS, \citeads{1999Msngr..95....8K}), FI (FIES, \citeads{2014AN....335...41T}), L (LAMOST, \citeads{2012RAA....12.1197C}), M (MMT, \citeads{2014ApJ...787...89B}), S (SDSS, \citeads{2013AJ....146...32S}), and U (UVES, \citeads{2000SPIE.4008..534D}). The approximate resolving power of the individual spectra, $\lambda/\Delta \lambda$ or $\Delta \lambda$, is given in brackets.} \tablefoottext{b}{The references for the atmospheric and stellar parameters are abbreviated as follows: C (current work), H \citepads{2017ApJ...847L...9H}, I \citepads{2019A&A...628L...5I}, KO \citepads{2020MNRAS.491.2465K}, KR \citepads{2020A&A...637A..53K}, and T \citepads{2009A&A...507L..37T}.}}
\end{table*}
\end{landscape}
\twocolumn
}
\begin{table*}
\vspace{-7pt}
\small
\centering
\setlength{\tabcolsep}{2.2pt}
\caption{\label{table:kinematics_AS}Kinematic parameters of the program stars for Galactic mass model~I of \citetads{2013A&A...549A.137I}.}
\vspace{-5.25pt}
\begin{tabular}{rrrrrrrrrrrrrrrrrrrrrrr}
\hline\hline
\# & $x$ & $y$ & $z$ & $r$ & & $\varv_x$ & $\varv_y$ & $\varv_z$ & $\varv_{\mathrm{Grf}}$ & $\varv_{\mathrm{Grf}}-\varv_{\mathrm{esc}}$ & $P_{\mathrm{b}}$& $x_{\mathrm{p}}$ & $y_{\mathrm{p}}$ & $z_{\mathrm{p}}$ & $r_{\mathrm{p}}$ & & $\varv_{x\mathrm{,p}}$ & $\varv_{y\mathrm{,p}}$& $\varv_{z\mathrm{,p}}$ & $\varv_{\mathrm{Grf,p}}$ & $\varv_{\mathrm{ej}}$ & $\tau_{\mathrm{flight}}$ \\\cline{2-5} \cline{7-11} \cline{13-16} \cline{18-22}& \multicolumn{4}{c}{(kpc)} & & \multicolumn{5}{c}{(km\,s$^{-1}$)} & (\%) & \multicolumn{4}{c}{(kpc)} & & \multicolumn{5}{c}{(km\,s$^{-1}$)} & (Myr) \\
\hline
1 & $-4.12$ & $-1.78$ & $-7.25$ & $8.52$ &   & $-720$ & $-353$ & $-1482$ & $1680$ & $1080$ & $0$ & $-0.6$ & $-0.06$ & $0.0$ & $0.19$ &   & $-790$ & $-364$ & $-1540$ & $1760$ & $1810$ & $4.72$ \\
  & $^{+0.17}_{-0.15}$ & $^{+0.06}_{-0.07}$ & $^{+0.24}_{-0.28}$ & $^{+0.18}_{-0.15}$ &   & $^{+50}_{-50}$ & $^{+14}_{-15}$ & $^{+21}_{-25}$ & $^{+50}_{-40}$ & $^{+50}_{-40}$ & \ldots & $^{+0.5}_{-0.4}$ & $^{+0.05}_{-0.05}$ & $^{+0.1}_{-0.2}$ & $^{+0.64}_{-0.09}$ &   & $^{+50}_{-60}$ & $^{+17}_{-24}$ & $^{+40}_{-50}$ & $^{+80}_{-60}$ & $^{+60}_{-60}$ & $^{+0.12}_{-0.11}$ \\
2 & $-21.6$ & $27.2$ & $-25.9$ & $43$ &   & $-203$ & $200$ & $-173$ & $334$ & $-124$ & $100$ & $2.3$ & $-1.2$ & $0.0$ & $3.4$ &   & $-180$ & $370$ & $-472$ & $600$ & $750$ & $105$ \\
  & $^{+1.4}_{-2.3}$ & $^{+4.6}_{-2.9}$ & $^{+2.8}_{-4.4}$ & $^{+7}_{-5}$ &   & $^{+28}_{-32}$ & $^{+23}_{-27}$ & $^{+16}_{-17}$ & $^{+15}_{-13}$ & $^{+24}_{-21}$ & \ldots & $^{+4.1}_{-1.9}$ & $^{+2.8}_{-2.0}$ & $^{+0.1}_{-0.1}$ & $^{+3.3}_{-1.9}$ &   & $^{+18}_{-42}$ & $^{+50}_{-60}$ & $^{+112}_{-\phantom{0}23}$ & $^{+50}_{-60}$ & $^{+60}_{-60}$ & $^{+16}_{-11}$ \\
3 & $-6.05$ & $5.1$ & $17.7$ & $19.3$ &   & $-108$ & $65$ & $305$ & $329$ & $-204$ & $100$ & $-0.1$ & $0.7$ & $0.0$ & $0.8$ &   & $-170$ & $210$ & $574$ & $640$ & $710$ & $44.0$ \\
  & $^{+0.22}_{-0.19}$ & $^{+0.5}_{-0.4}$ & $^{+1.7}_{-1.4}$ & $^{+1.6}_{-1.3}$ &   & $^{+13}_{-14}$ & $^{+20}_{-24}$ & $^{+9}_{-8}$ & $^{+9}_{-7}$ & $^{+16}_{-13}$ & \ldots & $^{+0.9}_{-0.7}$ & $^{+1.1}_{-0.7}$ & $^{+0.2}_{-0.1}$ & $^{+0.9}_{-0.5}$ &   & $^{+\phantom{0}50}_{-100}$ & $^{+40}_{-40}$ & $^{+30}_{-27}$ & $^{+50}_{-50}$ & $^{+70}_{-40}$ & $^{+3.4}_{-2.6}$ \\
4 & $1.8$ & $-0.51$ & $20.6$ & $20.7$ &   & $38$ & $0$ & $303$ & $305$ & $-222$ & $100$ & $-0.2$ & $-0.1$ & $0.0$ & $1.0$ &   & $10$ & $-20$ & $610$ & $610$ & $640$ & $50$ \\
  & $^{+0.9}_{-0.9}$ & $^{+0.05}_{-0.05}$ & $^{+1.9}_{-1.8}$ & $^{+1.9}_{-1.9}$ &   & $^{+24}_{-25}$ & $^{+40}_{-40}$ & $^{+13}_{-13}$ & $^{+11}_{-\phantom{0}9}$ & $^{+17}_{-13}$ & \ldots & $^{+0.9}_{-1.4}$ & $^{+1.2}_{-1.2}$ & $^{+0.2}_{-0.1}$ & $^{+1.0}_{-0.7}$ &   & $^{+110}_{-\phantom{0}60}$ & $^{+80}_{-60}$ & $^{+60}_{-50}$ & $^{+60}_{-50}$ & $^{+70}_{-50}$ & $^{+5}_{-5}$ \\
5 & $-32.4$ & $16.1$ & $23.2$ & $43$ &   & $-391$ & $335$ & $399$ & $652$ & $192$ & $0$ & $-8.9$ & $-3.0$ & $0.0$ & $9.5$ &   & $-502$ & $346$ & $453$ & $760$ & $627$ & $54$ \\
  & $^{+2.6}_{-2.7}$ & $^{+1.7}_{-1.8}$ & $^{+2.5}_{-2.6}$ & $^{+4}_{-4}$ &   & $^{+15}_{-15}$ & $^{+24}_{-25}$ & $^{+14}_{-12}$ & $^{+7}_{-7}$ & $^{+10}_{-10}$ & \ldots & $^{+1.2}_{-1.2}$ & $^{+1.5}_{-1.3}$ & $^{+0.1}_{-0.2}$ & $^{+1.1}_{-1.2}$ &   & $^{+20}_{-23}$ & $^{+17}_{-16}$ & $^{+11}_{-11}$ & $^{+12}_{-10}$ & $^{+34}_{-24}$ & $^{+5}_{-6}$ \\
6 & $-20.0$ & $-22.9$ & $44$ & $53$ &   & $-120$ & $10$ & $550$ & $563$ & $126$ & $0$ & $-11$ & $-22$ & $0$ & $25$ &   & $-170$ & $-50$ & $598$ & $626$ & $623$ & $75$ \\
  & $^{+1.0}_{-1.6}$ & $^{+1.9}_{-3.1}$ & $^{+6}_{-4}$ & $^{+7}_{-5}$ &   & $^{+80}_{-80}$ & $^{+60}_{-50}$ & $^{+40}_{-40}$ & $^{+31}_{-26}$ & $^{+34}_{-28}$ & \ldots & $^{+7}_{-6}$ & $^{+5}_{-6}$ & $^{+6}_{-5}$ & $^{+6}_{-6}$ &   & $^{+70}_{-70}$ & $^{+60}_{-60}$ & $^{+26}_{-25}$ & $^{+17}_{-14}$ & $^{+27}_{-26}$ & $^{+13}_{-12}$ \\
7 & $-6.10$ & $29.2$ & $-37$ & $48$ &   & $-210$ & $0$ & $-380$ & $430$ & $-20$ & $60$ & $11$ & $26$ & $0$ & $29$ &   & $-190$ & $90$ & $-428$ & $476$ & $612$ & $86$ \\
  & $^{+0.23}_{-0.32}$ & $^{+2.7}_{-4.1}$ & $^{+6}_{-4}$ & $^{+5}_{-7}$ &   & $^{+40}_{-40}$ & $^{+50}_{-50}$ & $^{+40}_{-40}$ & $^{+50}_{-50}$ & $^{+60}_{-70}$ & \ldots & $^{+5}_{-5}$ & $^{+7}_{-9}$ & $^{+5}_{-5}$ & $^{+\phantom{0}7}_{-10}$ &   & $^{+40}_{-40}$ & $^{+70}_{-60}$ & $^{+27}_{-26}$ & $^{+28}_{-16}$ & $^{+24}_{-21}$ & $^{+13}_{-13}$ \\
8 & $-18.1$ & $-8.4$ & $9.1$ & $21.9$ &   & $-523$ & $-124$ & $148$ & $557$ & $34$ & $2$ & $6.0$ & $-1.0$ & $0.0$ & $5.8$ &   & $-517$ & $-262$ & $327$ & $660$ & $589$ & $42$ \\
  & $^{+1.0}_{-1.0}$ & $^{+0.9}_{-0.8}$ & $^{+0.9}_{-0.9}$ & $^{+1.5}_{-1.5}$ &   & $^{+16}_{-16}$ & $^{+7}_{-7}$ & $^{+21}_{-21}$ & $^{+12}_{-11}$ & $^{+17}_{-17}$ & \ldots & $^{+3.5}_{-2.8}$ & $^{+0.8}_{-0.6}$ & $^{+0.1}_{-0.2}$ & $^{+3.3}_{-2.5}$ &   & $^{+10}_{-20}$ & $^{+13}_{-15}$ & $^{+\phantom{0}6}_{-20}$ & $^{+31}_{-23}$ & $^{+10}_{-13}$ & $^{+7}_{-6}$ \\
9 & $-15.9$ & $-22.1$ & $32.6$ & $42.5$ &   & $40$ & $-300$ & $184$ & $349$ & $-110$ & $100$ & $-14$ & $17$ & $0$ & $22$ &   & $-90$ & $-278$ & $310$ & $429$ & $586$ & $127$ \\
  & $^{+0.5}_{-0.9}$ & $^{+1.5}_{-2.4}$ & $^{+3.5}_{-2.1}$ & $^{+4.2}_{-2.6}$ &   & $^{+40}_{-40}$ & $^{+40}_{-40}$ & $^{+23}_{-26}$ & $^{+26}_{-18}$ & $^{+32}_{-24}$ & \ldots & $^{+6}_{-6}$ & $^{+11}_{-\phantom{0}8}$ & $^{+5}_{-6}$ & $^{+10}_{-\phantom{0}6}$ &   & $^{+50}_{-50}$ & $^{+33}_{-30}$ & $^{+22}_{-24}$ & $^{+11}_{-18}$ & $^{+17}_{-35}$ & $^{+27}_{-23}$ \\
10 & $-0.6$ & $13.6$ & $16$ & $21$ &   & $124$ & $260$ & $430$ & $514$ & $-14$ & $66$ & $-4.5$ & $2.7$ & $0.0$ & $5.30$ &   & $64$ & $385$ & $528$ & $642$ & $575$ & $34$ \\
  & $^{+1.6}_{-1.7}$ & $^{+2.8}_{-2.8}$ & $^{+4}_{-4}$ & $^{+5}_{-5}$ &   & $^{+5}_{-5}$ & $^{+50}_{-40}$ & $^{+40}_{-40}$ & $^{+12}_{-11}$ & $^{+30}_{-27}$ & \ldots & $^{+0.9}_{-0.8}$ & $^{+3.0}_{-2.3}$ & $^{+0.1}_{-0.1}$ & $^{+1.04}_{-0.25}$ &   & $^{+25}_{-38}$ & $^{+10}_{-33}$ & $^{+19}_{-26}$ & $^{+10}_{-11}$ & $^{+28}_{-47}$ & $^{+5}_{-5}$ \\
11 & $4.4$ & $4.31$ & $10.5$ & $12.1$ &   & $97$ & $272$ & $152$ & $326$ & $-247$ & $100$ & $-0.78$ & $-6.1$ & $0.0$ & $6.2$ &   & $148$ & $163$ & $380$ & $440$ & $561$ & $39.1$ \\
  & $^{+1.1}_{-0.7}$ & $^{+0.34}_{-0.22}$ & $^{+0.9}_{-0.6}$ & $^{+1.2}_{-0.8}$ &   & $^{+4}_{-4}$ & $^{+4}_{-4}$ & $^{+5}_{-5}$ & $^{+4}_{-4}$ & $^{+8}_{-7}$ & \ldots & $^{+0.36}_{-0.28}$ & $^{+0.4}_{-0.5}$ & $^{+0.1}_{-0.1}$ & $^{+0.5}_{-0.4}$ &   & $^{+19}_{-12}$ & $^{+5}_{-5}$ & $^{+5}_{-6}$ & $^{+4}_{-4}$ & $^{+15}_{-11}$ & $^{+3.2}_{-2.1}$ \\
12 & $-11.09$ & $-24.8$ & $39.9$ & $48$ &   & $-40$ & $-25$ & $446$ & $449$ & $1$ & $42$ & $-7$ & $-20$ & $0$ & $22$ &   & $-80$ & $-110$ & $506$ & $529$ & $561$ & $83$ \\
  & $^{+0.21}_{-0.31}$ & $^{+1.9}_{-2.9}$ & $^{+4.6}_{-3.0}$ & $^{+6}_{-4}$ &   & $^{+50}_{-50}$ & $^{+25}_{-26}$ & $^{+16}_{-17}$ & $^{+15}_{-14}$ & $^{+18}_{-16}$ & \ldots & $^{+4}_{-4}$ & $^{+4}_{-4}$ & $^{+5}_{-5}$ & $^{+4}_{-4}$ &   & $^{+40}_{-40}$ & $^{+40}_{-40}$ & $^{+12}_{-11}$ & $^{+9}_{-9}$ & $^{+16}_{-18}$ & $^{+13}_{-13}$ \\
13 & $-20.2$ & $63$ & $-20.8$ & $70$ &   & $-31$ & $532$ & $-205$ & $572$ & $162$ & $0$ & $-15.4$ & $9$ & $0.0$ & $16$ &   & $-92$ & $634$ & $-224$ & $673$ & $541$ & $93$ \\
  & $^{+1.1}_{-1.6}$ & $^{+9}_{-6}$ & $^{+1.9}_{-2.8}$ & $^{+9}_{-7}$ &   & $^{+18}_{-17}$ & $^{+10}_{-10}$ & $^{+23}_{-26}$ & $^{+6}_{-6}$ & $^{+14}_{-13}$ & \ldots & $^{+2.3}_{-2.5}$ & $^{+8}_{-7}$ & $^{+0.2}_{-0.1}$ & $^{+6}_{-4}$ &   & $^{+27}_{-33}$ & $^{+10}_{-24}$ & $^{+\phantom{0}9}_{-20}$ & $^{+15}_{-16}$ & $^{+26}_{-47}$ & $^{+11}_{-\phantom{0}9}$ \\
14 & $-14.6$ & $6.0$ & $-6.6$ & $17.2$ &   & $-10$ & $295$ & $-470$ & $551$ & $10$ & $40$ & $-14.1$ & $1.8$ & $0.0$ & $14.2$ &   & $-50$ & $307$ & $-490$ & $570$ & $500$ & $13.7$ \\
  & $^{+1.2}_{-1.2}$ & $^{+1.1}_{-1.2}$ & $^{+1.3}_{-1.2}$ & $^{+1.8}_{-1.9}$ &   & $^{+40}_{-40}$ & $^{+26}_{-26}$ & $^{+60}_{-60}$ & $^{+33}_{-27}$ & $^{+50}_{-40}$ & \ldots & $^{+1.5}_{-1.6}$ & $^{+1.2}_{-1.1}$ & $^{+0.2}_{-0.1}$ & $^{+1.8}_{-1.6}$ &   & $^{+40}_{-40}$ & $^{+24}_{-25}$ & $^{+60}_{-60}$ & $^{+30}_{-25}$ & $^{+50}_{-50}$ & $^{+0.9}_{-1.2}$ \\
15 & $-4.92$ & $4.0$ & $10.8$ & $12.5$ &   & $262$ & $149$ & $356$ & $466$ & $-105$ & $100$ & $-10.8$ & $-0.5$ & $0.0$ & $10.8$ &   & $161$ & $169$ & $432$ & $491$ & $470$ & $26.3$ \\
  & $^{+0.30}_{-0.30}$ & $^{+0.4}_{-0.4}$ & $^{+0.9}_{-0.9}$ & $^{+0.8}_{-0.8}$ &   & $^{+13}_{-13}$ & $^{+20}_{-18}$ & $^{+4}_{-4}$ & $^{+5}_{-5}$ & $^{+10}_{-\phantom{0}9}$ & \ldots & $^{+0.5}_{-0.6}$ & $^{+0.5}_{-0.5}$ & $^{+0.1}_{-0.2}$ & $^{+0.5}_{-0.5}$ &   & $^{+14}_{-13}$ & $^{+15}_{-15}$ & $^{+6}_{-7}$ & $^{+6}_{-5}$ & $^{+8}_{-9}$ & $^{+2.0}_{-2.0}$ \\
16 & $1.4$ & $0.43$ & $20.8$ & $20.9$ &   & $327$ & $80$ & $158$ & $371$ & $-155$ & $100$ & $-22$ & $-5.2$ & $0$ & $22$ &   & $200$ & $46$ & $297$ & $370$ & $436$ & $82$ \\
  & $^{+0.9}_{-0.9}$ & $^{+0.04}_{-0.04}$ & $^{+1.8}_{-1.9}$ & $^{+1.9}_{-1.9}$ &   & $^{+20}_{-19}$ & $^{+19}_{-19}$ & $^{+10}_{-10}$ & $^{+12}_{-11}$ & $^{+19}_{-18}$ & \ldots & $^{+5}_{-6}$ & $^{+1.3}_{-1.3}$ & $^{+5}_{-6}$ & $^{+5}_{-5}$ &   & $^{+40}_{-40}$ & $^{+15}_{-12}$ & $^{+11}_{-12}$ & $^{+17}_{-22}$ & $^{+\phantom{0}6}_{-22}$ & $^{+20}_{-19}$ \\
17 & $-0.9$ & $25.3$ & $23.1$ & $34.3$ &   & $165$ & $275$ & $321$ & $455$ & $-27$ & $97$ & $-10.3$ & $4.4$ & $0.0$ & $11.2$ &   & $91$ & $390$ & $403$ & $563$ & $436$ & $62$ \\
  & $^{+0.7}_{-0.7}$ & $^{+2.2}_{-2.2}$ & $^{+2.0}_{-2.0}$ & $^{+2.9}_{-3.0}$ &   & $^{+18}_{-17}$ & $^{+19}_{-18}$ & $^{+18}_{-17}$ & $^{+7}_{-6}$ & $^{+14}_{-13}$ & \ldots & $^{+1.1}_{-1.3}$ & $^{+2.5}_{-2.1}$ & $^{+0.1}_{-0.1}$ & $^{+1.9}_{-1.6}$ &   & $^{+27}_{-29}$ & $^{+16}_{-21}$ & $^{+9}_{-9}$ & $^{+8}_{-7}$ & $^{+12}_{-\phantom{0}9}$ & $^{+5}_{-4}$ \\
18 & $-29.5$ & $-13.0$ & $26.1$ & $41.5$ &   & $-380$ & $18$ & $261$ & $460$ & $-5$ & $55$ & $5$ & $-10.7$ & $0.0$ & $11.6$ &   & $-430$ & $-120$ & $367$ & $575$ & $427$ & $82$ \\
  & $^{+1.8}_{-1.8}$ & $^{+1.1}_{-1.1}$ & $^{+2.1}_{-2.3}$ & $^{+3.0}_{-3.0}$ &   & $^{+27}_{-26}$ & $^{+22}_{-23}$ & $^{+19}_{-19}$ & $^{+16}_{-14}$ & $^{+19}_{-16}$ & \ldots & $^{+5}_{-4}$ & $^{+1.8}_{-1.9}$ & $^{+0.1}_{-0.1}$ & $^{+3.3}_{-2.1}$ &   & $^{+10}_{-10}$ & $^{+29}_{-29}$ & $^{+11}_{-15}$ & $^{+9}_{-8}$ & $^{+13}_{-\phantom{0}7}$ & $^{+10}_{-\phantom{0}9}$ \\
19 & $-26.1$ & $-5.7$ & $26.2$ & $37.4$ &   & $-328$ & $145$ & $273$ & $451$ & $-22$ & $93$ & $3.7$ & $-14.8$ & $0.0$ & $15.1$ &   & $-390$ & $23$ & $374$ & $540$ & $417$ & $80$ \\
  & $^{+1.5}_{-2.4}$ & $^{+0.5}_{-0.8}$ & $^{+3.5}_{-2.2}$ & $^{+4.3}_{-2.6}$ &   & $^{+14}_{-16}$ & $^{+\phantom{0}9}_{-10}$ & $^{+10}_{-12}$ & $^{+7}_{-6}$ & $^{+14}_{-11}$ & \ldots & $^{+3.3}_{-2.3}$ & $^{+1.3}_{-1.6}$ & $^{+0.1}_{-0.1}$ & $^{+2.5}_{-1.7}$ &   & $^{+4}_{-4}$ & $^{+13}_{-13}$ & $^{+5}_{-6}$ & $^{+4}_{-5}$ & $^{+9}_{-6}$ & $^{+13}_{-\phantom{0}8}$ \\
20 & $-7.92$ & $0.439$ & $-1.51$ & $8.08$ &   & $-370$ & $389$ & $-183$ & $567$ & $-50$ & $100$ & $-4.8$ & $-2.55$ & $0.0$ & $5.44$ &   & $-435$ & $375$ & $-203$ & $609$ & $414$ & $7.6$ \\
  & $^{+0.06}_{-0.06}$ & $^{+0.021}_{-0.022}$ & $^{+0.08}_{-0.08}$ & $^{+0.05}_{-0.05}$ &   & $^{+20}_{-20}$ & $^{+6}_{-6}$ & $^{+6}_{-5}$ & $^{+19}_{-19}$ & $^{+18}_{-19}$ & \ldots & $^{+0.4}_{-0.4}$ & $^{+0.20}_{-0.21}$ & $^{+0.1}_{-0.2}$ & $^{+0.23}_{-0.21}$ &   & $^{+23}_{-22}$ & $^{+6}_{-6}$ & $^{+7}_{-7}$ & $^{+21}_{-21}$ & $^{+18}_{-19}$ & $^{+0.6}_{-0.5}$ \\
21 & $-5.82$ & $2.26$ & $-5.5$ & $8.32$ &   & $-71$ & $98$ & $-165$ & $204$ & $-404$ & $100$ & $-2.4$ & $-0.42$ & $0.0$ & $2.5$ &   & $-266$ & $119.9$ & $-318$ & $431$ & $402$ & $23.1$ \\
  & $^{+0.17}_{-0.17}$ & $^{+0.15}_{-0.14}$ & $^{+0.4}_{-0.4}$ & $^{+0.17}_{-0.15}$ &   & $^{+7}_{-7}$ & $^{+12}_{-11}$ & $^{+8}_{-8}$ & $^{+4}_{-4}$ & $^{+6}_{-5}$ & \ldots & $^{+0.4}_{-0.4}$ & $^{+0.28}_{-0.29}$ & $^{+0.1}_{-0.1}$ & $^{+0.4}_{-0.4}$ &   & $^{+11}_{-11}$ & $^{+2.3}_{-2.1}$ & $^{+19}_{-21}$ & $^{+22}_{-20}$ & $^{+33}_{-30}$ & $^{+0.5}_{-0.5}$ \\
22 & $-9.11$ & $0.499$ & $3.23$ & $9.68$ &   & $-336$ & $395$ & $183$ & $549$ & $-49$ & $100$ & $-3.1$ & $-5.5$ & $0.0$ & $6.28$ &   & $-435$ & $343$ & $229.6$ & $599$ & $388$ & $15.3$ \\
  & $^{+0.07}_{-0.06}$ & $^{+0.025}_{-0.024}$ & $^{+0.16}_{-0.15}$ & $^{+0.10}_{-0.10}$ &   & $^{+15}_{-14}$ & $^{+6}_{-6}$ & $^{+4}_{-4}$ & $^{+12}_{-11}$ & $^{+12}_{-12}$ & \ldots & $^{+0.6}_{-0.6}$ & $^{+0.4}_{-0.4}$ & $^{+0.2}_{-0.1}$ & $^{+0.08}_{-0.07}$ &   & $^{+17}_{-16}$ & $^{+5}_{-6}$ & $^{+1.2}_{-1.1}$ & $^{+10}_{-11}$ & $^{+15}_{-13}$ & $^{+1.0}_{-1.0}$ \\
23 & $-19.2$ & $-8.3$ & $-10.0$ & $23.1$ &   & $-141.7$ & $-4$ & $30$ & $145$ & $-374$ & $100$ & $6.0$ & $0.3$ & $0.0$ & $6.0$ &   & $-157$ & $-189$ & $-331$ & $414$ & $381$ & $109$ \\
  & $^{+0.5}_{-0.7}$ & $^{+0.4}_{-0.5}$ & $^{+0.5}_{-0.6}$ & $^{+1.0}_{-0.7}$ &   & $^{+2.5}_{-2.4}$ & $^{+\phantom{0}8}_{-10}$ & $^{+9}_{-7}$ & $^{+4}_{-4}$ & $^{+8}_{-6}$ & \ldots & $^{+0.9}_{-0.7}$ & $^{+0.7}_{-0.6}$ & $^{+0.2}_{-0.1}$ & $^{+1.0}_{-0.7}$ &   & $^{+13}_{-14}$ & $^{+27}_{-19}$ & $^{+6}_{-5}$ & $^{+12}_{-16}$ & $^{+10}_{-12}$ & $^{+8}_{-6}$ \\
24 & $-5.55$ & $1.76$ & $-5.8$ & $8.20$ &   & $41.6$ & $51$ & $-207$ & $218$ & $-392$ & $100$ & $-5.08$ & $0.3$ & $0.0$ & $5.07$ &   & $-111$ & $76$ & $-311$ & $339$ & $374$ & $21.7$ \\
  & $^{+0.18}_{-0.18}$ & $^{+0.11}_{-0.11}$ & $^{+0.4}_{-0.4}$ & $^{+0.17}_{-0.15}$ &   & $^{+2.9}_{-3.2}$ & $^{+15}_{-15}$ & $^{+7}_{-7}$ & $^{+4}_{-4}$ & $^{+6}_{-5}$ & \ldots & $^{+0.23}_{-0.23}$ & $^{+0.4}_{-0.4}$ & $^{+0.1}_{-0.1}$ & $^{+0.22}_{-0.19}$ &   & $^{+6}_{-7}$ & $^{+8}_{-7}$ & $^{+12}_{-13}$ & $^{+11}_{-11}$ & $^{+19}_{-19}$ & $^{+0.8}_{-0.8}$ \\
25 & $-26.7$ & $4.9$ & $-11.5$ & $29.5$ &   & $-336$ & $213$ & $-153$ & $426$ & $-70$ & $100$ & $-2.1$ & $-7.5$ & $0.0$ & $7.9$ &   & $-502$ & $130$ & $-235$ & $568$ & $362$ & $60$ \\
  & $^{+1.8}_{-1.8}$ & $^{+0.5}_{-0.5}$ & $^{+1.2}_{-1.1}$ & $^{+2.0}_{-2.2}$ &   & $^{+8}_{-9}$ & $^{+14}_{-14}$ & $^{+8}_{-8}$ & $^{+6}_{-6}$ & $^{+7}_{-7}$ & \ldots & $^{+1.6}_{-1.3}$ & $^{+0.5}_{-0.5}$ & $^{+0.1}_{-0.1}$ & $^{+0.5}_{-0.6}$ &   & $^{+13}_{-11}$ & $^{+26}_{-27}$ & $^{+\phantom{0}8}_{-11}$ & $^{+10}_{-\phantom{0}9}$ & $^{+13}_{-10}$ & $^{+6}_{-6}$ \\
26 & $-31.0$ & $8.0$ & $16.3$ & $35.9$ &   & $-207$ & $140$ & $147$ & $290$ & $-187$ & $100$ & $-6.1$ & $-4.9$ & $0.0$ & $8.1$ &   & $-416$ & $115$ & $238$ & $488$ & $360$ & $86$ \\
  & $^{+2.4}_{-2.3}$ & $^{+0.9}_{-0.9}$ & $^{+1.7}_{-1.7}$ & $^{+3.0}_{-3.0}$ &   & $^{+\phantom{0}9}_{-10}$ & $^{+21}_{-22}$ & $^{+10}_{-10}$ & $^{+8}_{-7}$ & $^{+8}_{-7}$ & \ldots & $^{+1.7}_{-1.5}$ & $^{+1.6}_{-1.3}$ & $^{+0.1}_{-0.1}$ & $^{+1.2}_{-1.6}$ &   & $^{+25}_{-29}$ & $^{+13}_{-18}$ & $^{+12}_{-11}$ & $^{+29}_{-20}$ & $^{+50}_{-40}$ & $^{+9}_{-8}$ \\
27 & $-6.42$ & $-16.9$ & $-9.7$ & $20.6$ &   & $-381$ & $-161$ & $-242$ & $479$ & $-50$ & $98$ & $7.5$ & $-9.2$ & $0.0$ & $11.8$ &   & $-361$ & $-274$ & $-287.7$ & $537$ & $355$ & $35.5$ \\
  & $^{+0.12}_{-0.12}$ & $^{+0.9}_{-0.9}$ & $^{+0.5}_{-0.5}$ & $^{+0.9}_{-1.0}$ &   & $^{+24}_{-22}$ & $^{+4}_{-4}$ & $^{+4}_{-4}$ & $^{+19}_{-20}$ & $^{+23}_{-24}$ & \ldots & $^{+1.5}_{-1.6}$ & $^{+0.5}_{-0.6}$ & $^{+0.2}_{-0.1}$ & $^{+1.4}_{-1.3}$ &   & $^{+17}_{-18}$ & $^{+5}_{-5}$ & $^{+2.3}_{-2.2}$ & $^{+11}_{-10}$ & $^{+13}_{-\phantom{0}4}$ & $^{+1.8}_{-1.8}$ \\
28 & $1.0$ & $-2.62$ & $-6.1$ & $6.7$ &   & $-286$ & $270.2$ & $-75$ & $400$ & $-224$ & $100$ & $10.3$ & $-10.6$ & $0.0$ & $14.8$ &   & $-179$ & $139$ & $-188.7$ & $295$ & $355$ & $38.6$ \\
  & $^{+0.5}_{-0.6}$ & $^{+0.14}_{-0.14}$ & $^{+0.4}_{-0.4}$ & $^{+0.5}_{-0.4}$ &   & $^{+8}_{-7}$ & $^{+2.8}_{-2.9}$ & $^{+11}_{-10}$ & $^{+7}_{-6}$ & $^{+11}_{-11}$ & \ldots & $^{+0.9}_{-0.9}$ & $^{+0.5}_{-0.5}$ & $^{+0.1}_{-0.1}$ & $^{+0.9}_{-1.0}$ &   & $^{+4}_{-5}$ & $^{+8}_{-9}$ & $^{+2.0}_{-2.3}$ & $^{+8}_{-8}$ & $^{+10}_{-11}$ & $^{+1.1}_{-1.1}$ \\
29 & $-6.66$ & $-0.85$ & $2.86$ & $7.30$ &   & $-98$ & $463$ & $166$ & $502$ & $-122$ & $100$ & $-4.52$ & $-7.4$ & $0.0$ & $8.69$ &   & $-181$ & $404$ & $209$ & $489$ & $349$ & $14.6$ \\
  & $^{+0.10}_{-0.10}$ & $^{+0.05}_{-0.05}$ & $^{+0.14}_{-0.15}$ & $^{+0.05}_{-0.05}$ &   & $^{+7}_{-7}$ & $^{+11}_{-11}$ & $^{+7}_{-7}$ & $^{+14}_{-14}$ & $^{+14}_{-13}$ & \ldots & $^{+0.23}_{-0.21}$ & $^{+0.4}_{-0.4}$ & $^{+0.1}_{-0.1}$ & $^{+0.17}_{-0.18}$ &   & $^{+5}_{-5}$ & $^{+10}_{-\phantom{0}9}$ & $^{+8}_{-9}$ & $^{+13}_{-13}$ & $^{+18}_{-17}$ & $^{+0.6}_{-0.5}$ \\
30 & $-8.69$ & $-1.46$ & $2.72$ & $9.22$ &   & $-60.1$ & $288$ & $315$ & $431$ & $-173$ & $100$ & $-7.98$ & $-3.82$ & $0.0$ & $8.85$ &   & $-109$ & $272$ & $330$ & $442$ & $335$ & $8.2$ \\
  & $^{+0.06}_{-0.06}$ & $^{+0.08}_{-0.08}$ & $^{+0.14}_{-0.14}$ & $^{+0.09}_{-0.08}$ &   & $^{+2.8}_{-2.8}$ & $^{+9}_{-8}$ & $^{+5}_{-5}$ & $^{+9}_{-9}$ & $^{+10}_{-\phantom{0}9}$ & \ldots & $^{+0.07}_{-0.08}$ & $^{+0.23}_{-0.23}$ & $^{+0.1}_{-0.1}$ & $^{+0.08}_{-0.08}$ &   & $^{+4}_{-5}$ & $^{+7}_{-7}$ & $^{+5}_{-5}$ & $^{+9}_{-9}$ & $^{+7}_{-6}$ & $^{+0.5}_{-0.5}$ \\
\hline
\end{tabular}
\vspace{-6pt}
\tablefoot{The identification numbers in the first column correspond to those in Table~\ref{table:vejection}. Results and statistical uncertainties are derived with a Monte Carlo simulation and are either given as the mode and highest density interval of $1\sigma$ confidence if the resulting parameter distribution is unimodal or as the median value plus 15.87th and 84.13th percentiles if it is not. In addition to Cartesian positions and velocities, the Galactic rest-frame velocity $\varv_{\mathrm{Grf}}=(\varv_x^2+\varv_y^2+\varv_z^2)^{1/2}$, the local Galactic escape velocity $\varv_{\mathrm{esc}}$, the galactocentric radius $r=(x^2+y^2+z^2)^{1/2}$, the ejection velocity $\varv_{\mathrm{ej}}$ (defined as the Galactic rest-frame velocity relative to the rotating Galactic disk), and the flight time $\tau_{\mathrm{flight}}$ are listed. Plane-crossing quantities are labeled with the subscript ``p''. The probability $P_{\mathrm{b}}$ is the fraction of Monte Carlo runs for which the star is bound to the Milky Way.}
\end{table*}
\end{appendix}
\end{document}